\documentstyle[preprint,aps]{revtex}
\begin{document}
\tightenlines
\title{Nonlocal ghost formalism for quantum gravity}
\author{{Giampiero Esposito}
\thanks{Electronic address: giampiero.esposito@na.infn.it}}
\address{Istituto Nazionale di Fisica Nucleare, 
Sezione di Napoli,\\
Mostra d'Oltremare Padiglione 20, 80125 Napoli, Italy\\
Dipartimento di Scienze Fisiche,\\
Complesso Universitario di Monte S. Angelo,\\
Via Cintia, Edificio G, 80126 Napoli, Italy}
\maketitle
\begin{abstract}
The quantum theory of a free particle on a portion of 
two-dimensional Euclidean space bounded by a circle and subject
to non-local boundary conditions gives rise to bulk and surface
states. Starting from this well known property, a counterpart
for gravity is here considered. In particular, if spatial
components of metric perturbations are set to zero at the
boundary, invariance of the full set of boundary conditions 
under infinitesimal diffeomorphisms is compatible with non-local
boundary conditions on normal components of metric perturbations
if and only if both the gauge-field operator and the ghost
operator are pseudo-differential operators in one-loop quantum
gravity.
\end{abstract}
Work over the last twelve years on the problem of boundary conditions
in (one-loop) Euclidean quantum gravity has focused on a local
formulation, by trying to satisfy a number of basic requirements,
which are as follows [1].
\vskip 0.3cm
\noindent
(i) Local nature of the boundary operators.
\vskip 0.3cm
\noindent
(ii) Operator on metric perturbations and ghost operator of
Laplace type and essentially self-adjoint.
\vskip 0.3cm
\noindent
(iii) Strong ellipticity of the boundary-value problems for
such differential operators with local boundary conditions.
\vskip 0.3cm
\noindent
(iv) Gauge invariance of the boundary conditions and/or of
the out-in (one-loop) amplitude.
\vskip 0.3cm
\noindent
At about the same time, in papers devoted to quantum field
theory and quantum gravity, non-local boundary conditions had
been studied mainly for operators of Dirac type, with the
exception of the work in Ref. [2], where a set of non-local
boundary conditions for the quantized gravitational field
was analyzed in detail. On the other hand, non-local boundary 
conditions for operators of Laplace type had already been studied
quite intensively in the mathematical literature dealing with
functional calculus of pseudo-differential boundary-value problems [3].
Here, however, we are concerned with physical applications, and
hence we begin our presentation, following Ref. [4], with a problem
inspired by Bose--Einstein condensation models, i.e. the quantum
theory of a free particle in two dimensions subject to non-local
boundary conditions on a circle.

More precisely, given the function $q \in L_{1}({\bf R}) \cap
L_{2}({\bf R})$, one may define
\begin{equation}
q_{R}(x) \equiv {1\over 2\pi R} \sum_{l=-\infty}^{\infty}
e^{ilx/R} \int_{-\infty}^{\infty} e^{-ily/R} q(y) dy.
\label{1}
\end{equation}
The function $q_{R}$ is, by construction, periodic with period
$2\pi R$, and tends to $q$ as $R$ tends to $\infty$. 
On considering the region
\begin{equation}
B_{R} \equiv \left \{ x,y: x^{2}+y^{2} \leq R^{2} \right \},
\label{2}
\end{equation}
one studies the Laplacian acting on square-integrable functions
on $B_{R}$, with non-local boundary conditions given by [4]
\begin{equation}
[u_{;N}]_{\partial B_{R}}+ \oint_{\partial B_{R}}
q_{R}(s-s') u(R \cos(s'/R), R \sin(s'/R))ds'=0.
\label{3}
\end{equation}
In polar coordinates, the resulting boundary-value problem reads
\begin{equation}
-\left({\partial^{2}\over \partial r^{2}}+{1\over r}
{\partial \over \partial r}+{1\over r^{2}}
{\partial^{2} \over \partial \varphi^{2}} \right)u=Eu,
\label{4}
\end{equation}
\begin{equation}
{\partial u \over \partial r}(R,\varphi)
+R \int_{-\pi}^{\pi} q_{R}(R(\varphi-\theta))u(R,\theta)d\theta=0.
\label{5}
\end{equation}
For example, when the eigenvalue $E$ is positive in Eq. (4), the
corresponding eigenfunction reads
\begin{equation}
u_{l,E}(r,\varphi)=J_{l}(r \sqrt{E}) e^{il \varphi},
\label{6}
\end{equation}
where $J_{l}$ is the standard notation for the Bessel function of
first kind of order $l \in Z$. On denoting by $\widetilde q$ the
Fourier transform of $q$, and inserting the formula (6) into the
boundary condition (5), one finds an equation leading, implicitly,
to the knowledge of positive eigenvalues, i.e.
\begin{equation}
\Bigr[\sqrt{E}J_{l}'(R \sqrt{E})+J_{l}(R \sqrt{E})
{\widetilde q}(l/R) \Bigr]=0.
\label{7}
\end{equation}
The solutions which decay rapidly away from the boundary are the
{\it surface states}, whereas the solutions which remain 
non-negligible are called {\it bulk states} [4].

Since so many interesting new features arise already at the
level of non-relativistic quantum mechanics when non-local
boundary conditions are imposed, a naturally occurring question
is whether a counterpart exists in the quantum theory of the
gravitational field in the Euclidean regime. Indeed, the one-loop
quantum theory and the use of local boundary conditions are known
to lead to operators of Laplace type on metric perturbations, if
linear covariant gauges are used in the path integral for
out-in amplitudes. Let us also recall that in Euclidean quantum
gravity the request of invariance under infinitesimal diffeomorphisms
of homogeneous Dirichlet conditions on spatial components $h_{ij}$ of
metric perturbations leads to homogeneous Dirichlet conditions on
the whole ghost one-form [5--7] for all boundaries which are not
totally geodesic (a boundary is said to be totally geodesic when
its extrinsic curvature tensor vanishes). At that stage, the vanishing
of the gauge-averaging functional $\Phi_{a}(h)$ at the boundary
is imposed to ensure that this remaining set of boundary conditions
on metric perturbations leads again to homogeneous Dirichlet
conditions on the ghost (with the exception of zero-modes [8]). Here,
however, we would like to consider non-local boundary operators.

We therefore assume that $\Phi_{a}(h)$ consists of the de Donder
term (which has the advantage of leading to an operator of Laplace
type on metric perturbations with purely local boundary conditions)
plus a pseudo-differential part with suitably smooth kernel, i.e.
(here ${\hat h} \equiv g^{cd}h_{cd}$)
\begin{equation}
\Phi_{a}(h) \equiv \nabla^{b} 
\left(h_{ab}-{1\over 2}g_{ab}{\hat h} \right)
+(\zeta h)_{a},
\label{8}
\end{equation}
where 
\begin{equation}
(\zeta h)_{a}(x) \equiv \int_{M} \zeta_{a}^{\; cd}(x,x')
h_{cd}(x')dV'.
\label{9}
\end{equation}
As in Ref. [8], $\nabla$ is the connection on the bundle of symmetric
rank-2 tensor fields over $M$, $g$ is the background metric,
and $dV'$ is the integration measure over $M$. Moreover,
$\zeta_{a}^{\; cd}(x,x')$ is taken to be a  
kernel on the curved $m$-dimensional Riemannian background
$(M,g)$ (for technical details, see the appendix of Ref. [9]). 
By virtue of the assumption (9), the standard rule for
the evaluation of the ghost operator ${\cal F}_{a}^{\; b}$ yields
now a non-trivial result, because 
\begin{equation}
\Phi_{a}(h)-\Phi_{a}({ }^{\varepsilon}h)
={\cal F}_{a}^{\; b} \; \varepsilon_{b},
\label{10}
\end{equation}
where (here $\Box \equiv g^{ab}\nabla_{a}\nabla_{b}$)
\begin{equation}
{\cal F}_{a}^{\; b} \equiv -\delta_{a}^{\; b} \Box 
-R_{a}^{\; b}+{\widetilde {\cal F}}_{a}^{\; b},
\label{11}
\end{equation}
${\widetilde {\cal F}}_{a}^{\; b}$ being the pseudo-differential
operator defined by
\begin{equation}
{\widetilde {\cal F}}_{a}^{\; b} \; \varepsilon_{b}(x)
\equiv - \int_{M} \zeta_{a}^{\; cd}(x,x')\nabla_{(c} \; 
\varepsilon_{d)}(x')dV'.
\label{12}
\end{equation}

The gauge-field operator $P_{ab}^{\; \; \; cd}$ on metric 
perturbations is obtained by expanding to quadratic order in
$h_{ab}$ the Euclidean Einstein--Hilbert action and adding the 
integral over $M$ of ${\Phi_{a}(h)\Phi^{a}(h)\over 2\alpha}$.
On setting $\alpha=1$ for simplicity, one finds that
\begin{equation}
P_{ab}^{\; \; \; cd}=G_{ab}^{\; \; \; cd}
+U_{ab}^{\; \; \; cd}+V_{ab}^{\; \; \; cd}.
\label{13}
\end{equation}
With our notation, $G_{ab}^{\; \; \; cd}$ is the operator of
Laplace type given by [7]
\begin{eqnarray}
G_{ab}^{\; \; \; cd} & \equiv & E_{ab}^{\; \; \; cd}
\left(-\Box +R \right)
-2E_{ab}^{\; \; \; qf} \; R_{\; qpf}^{c} \; g^{dp} 
\nonumber \\
&-& E_{ab}^{\; \; \; pd} \; R_{p}^{\; c}
-E_{ab}^{\; \; \; cp} \; R_{p}^{\; d},
\label{14} 
\end{eqnarray}
where $E^{abcd}$ is the DeWitt supermetric (i.e. the metric on
the vector bundle of symmetric rank-2 tensor fields over $M$)
\begin{equation}
E^{abcd} \equiv {1\over 2}\Bigr(g^{ac}g^{bd}+g^{ad}g^{bc}
-g^{ab}g^{cd}\Bigr).
\label{15}
\end{equation}
Moreover, $U_{ab}^{\; \; \; cd}$ and $V_{ab}^{\; \; \; cd}$ are
pseudo-differential operators resulting from
$\Phi_{a}(h)(\zeta h)^{a}$ and ${(\zeta h)_{a}(\zeta h)^{a}\over 2}$
respectively, in the expression of the gauge-averaging term 
${\Phi_{a}(h)\Phi^{a}(h)\over 2}$. To work out the former, we
use the identity
\begin{eqnarray}
\nabla^{b}T_{b} & \equiv & 
\nabla^{b}\left[\left(h_{ab}-{1\over 2}g_{ab}{\hat h} \right)
\int_{M}\zeta^{acd}(x,x')h_{cd}(x')dV' \right] 
\nonumber \\
&=& \left[\nabla^{b}\left(h_{ab}-{1\over 2}g_{ab}{\hat h} \right)\right]
\int_{M}\zeta^{acd}(x,x')h_{cd}(x')dV' 
\nonumber \\
&+& \left(h_{ab}-{1\over 2}g_{ab}{\hat h} \right)\nabla^{b}
\int_{M}\zeta^{acd}(x,x')h_{cd}(x')dV'.
\label{16}
\end{eqnarray}
We therefore define
\begin{equation}
T_{b} \equiv \left(h_{ab}-{1\over 2}g_{ab}{\hat h}\right)
(\zeta h)^{a},
\label{17}
\end{equation}
and add to the action a boundary term equal to
($d\Sigma'$ being the integration measure over $\partial M$)
$$
-\int_{\partial M}N^{b}T_{b}d\Sigma'
$$
to find that the action of $U_{ab}^{\; \; \; cd}$ is defined by
\begin{equation}
U_{ab}^{\; \; \; cd}h_{cd}(x) \equiv -2 \nabla^{r}E_{rsab}
\int_{M} \zeta^{scd}(x,x')h_{cd}(x')dV'.
\label{18}
\end{equation}
Furthermore, the definition (9) implies that 
$V_{ab}^{\; \; \; cd}$ is a pseudo-differential operator whose 
action is defined by
\begin{eqnarray}
\; & \; & h^{ab} \; V_{ab}^{\; \; \; cd} \; h_{cd}(x) 
\nonumber \\
& \equiv & \int_{M^{2}} h^{ab}(x') \zeta_{pab}(x,x')
\zeta^{pcd}(x,x'')h_{cd}(x'')dV' dV''.
\label{19}
\end{eqnarray}

To sum up, we have imposed the following boundary
conditions on metric perturbations:
\begin{equation}
[h_{ij}]_{\partial M}=0,
\label{20}
\end{equation}
\begin{equation}
\Bigr[\Phi_{a}(h)\Bigr]_{\partial M}
=\left[\nabla^{b}\left(h_{ab}-{1\over 2}g_{ab}{\hat h} \right)
+(\zeta h)_{a}\right]_{\partial M}=0.
\label{21}
\end{equation}
Both (20) and (21) are invariant under infinitesimal 
diffeomorphisms if the whole ghost one-form vanishes
at the boundary, i.e.
\begin{equation}
[\varepsilon_{a}]_{\partial M}=0.
\label{22}
\end{equation}
Thus, we have proposed the general equations for an approach
to Euclidean quantum gravity where both metric perturbations and
ghost fields are ruled by pseudo-differential operators, while 
the boundary conditions have a Dirichlet and a pseudo-differential
sector (and are pure Dirichlet for the ghost). As we know from the
general analysis in Refs. [1] and [3], the adjoint of the ghost 
operator will instead be a differential operator, but subject to
non-local boundary conditions, to ensure self-adjointness.

Interestingly (at the risk of repeating ourselves), non-local
boundary conditions, jointly with the request of their complete
invariance under infinitesimal diffeomorphisms, 
are sufficient to lead to gauge-field and ghost
operators of pseudo-differential nature. On the mathematical side,
it remains to be seen under which restrictions the above boundary
conditions lead to strong ellipticity [3,8] in one-loop quantum
gravity. On the physical side, the applications to the wave 
function of the universe [10] are entirely unexplored and might
lead to a better understanding of quantum cosmology.

\acknowledgments
This work has been partially supported by PRIN97 ``Sintesi''.
We are indebted to Ivan Avramidi and Alexander Kamenshchik for
scientific collaboration on Euclidean quantum gravity.

\end{document}